\input{aipcheck}

\documentclass[
    ,final            
  ]
  {aipproc}

\layoutstyle{8x11single}

\begin{document}

\title{Interpreting the peak structures around 1800 MeV in the BES data on $J/\Psi \to \phi \pi^+ \pi^-$, $J/\Psi\to \gamma \omega \phi$}


\author{K. P. Khemchandani}{
  address={Instituto de F\'isica, Universidade de S\~ao Paulo, C.P 66318, 05314-970 S\~ao Paulo, SP, Brazil.}
}
\author{A. Mart\'inez Torres}{
  address={Instituto de F\'isica, Universidade de S\~ao Paulo, C.P 66318, 05314-970 S\~ao Paulo, SP, Brazil.}
}

\author{M. Nielsen}{
  address={Instituto de F\'isica, Universidade de S\~ao Paulo, C.P 66318, 05314-970 S\~ao Paulo, SP, Brazil.}
}

\author{F. S. Navarra}{
  address={Instituto de F\'isica, Universidade de S\~ao Paulo, C.P 66318, 05314-970 S\~ao Paulo, SP, Brazil.}
}

\author{D. Jido}{
  address={Department of Physics, Tokyo Metropolitan University, Hachioji, Tokyo 192-0397, Japan.}
}
\author{A. Hosaka}{
  address={Research Center for Nuclear Physics (RCNP), Mihogaoka 10-1, Ibaraki 567-0047, Japan.}
}
\author{E. Oset}{
  address={Departamento de F\' isica Te\' orica and IFIC, Centro Mixto Universidad de Valencia-CSIC, Institutos de Investigaci\'on
de Paterna, Aptd. 22085, 46071 Valencia, Spain.}
}

\begin{abstract}
In this talk we present an interpretation for the experimental data available on two different processes, namely,
$J/\Psi \to \phi \pi^+ \pi^-$, $J/\Psi\to \gamma \omega \phi$, which seem to indicate existence of two new resonances
with the same quantum numbers ($J^{\pi c}=0^{++}, I = 0$) and very similar mass (~1800 MeV) but with very different decay 
properties. However, our studies show that the peak structure found in the $\omega \phi$ invariant mass, in 
$J/\Psi \to \gamma \omega \phi$, is a manifestation of the well known $f_0(1710)$ while the cross section enhancement
found in $J/\Psi \to \phi \pi^+ \pi^-$ is indeed a new $f_0$ resonance with mass near 1800 MeV. We present an explanation for 
 the different decay properties  of these two scalar resonances.
 \end{abstract}

\maketitle


\section{Introduction}
A peak structure has been found in the $\phi \omega$ invariant mass spectrum in
 a recent experimental study of the $J/\Psi \to \gamma \phi \omega$ process  made by the BES
collaboration \cite{phiwbes}. This peak has been associated to a scalar, isoscalar resonance 
with mass $M = 1795 \pm 7^{+23}_{-5}$ MeV and width $\Gamma = 95 \pm10^{+78}_{-34}$ MeV  in Ref.~\cite{phiwbes}. We shall 
refer to this state as $f_0(1800)$ as done in Ref.~\cite{phiwbes}.
No such resonance is present in the spectrum of the $f_0$ states in the particle data book (PDB) \cite{pdg}. The known state with 
the nearest possible  mass  is $f_0(1710)$ and one could wonder if the peak seen in the  $\phi \omega$ spectrum can be
explained with this known resonance. In fact, an attempt to explain the BES data  on the $\phi \omega$ invariant mass
spectrum with $f_0(1710)$ was made in Ref.~\cite{zhou} and indeed a peak structure was found near the threshold
although with a strength much weaker than the experimental data. Anyhow, no strong claims were made
in Ref.~\cite{zhou} partly due to a parameterized  treatment of the interaction of vector mesons ($V$) and $V V f_0(1710)$ coupling.

Interestingly, the finding of yet another $f_0$ resonance with mass also around 1800 MeV was earlier reported in
a different process: $J/\Psi \to  \phi \pi^+ \pi^-$  \cite{pipibes}. This resonance was found in the $\pi \pi$ invariant mass
spectrum and was named $f_0(1790)$. This state was  found to possess  a very curious decay property which is a suppressed
decay to the $K \overline{K}$ channel.  This property makes $f_0(1790)$ undoubtedly distinct to the known $f_0(1710)$ which, completely contrarily, decays to 
$K \overline{K}$ with a large branching ratio and suppressively   to $\pi \pi$.  

\begin{figure}[h!]
  \includegraphics[height=2cm]{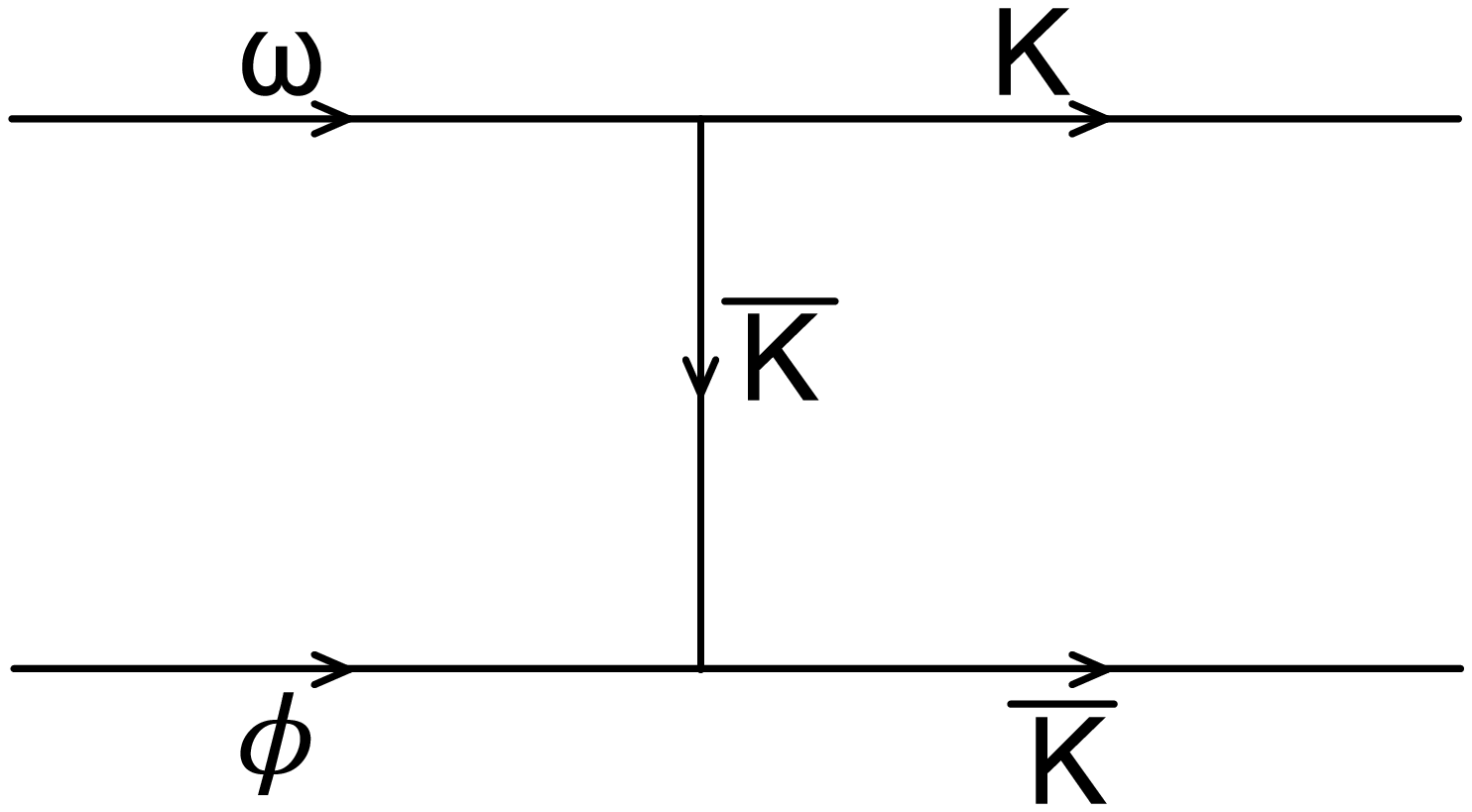}
\caption{The $\phi \omega \to K \overline{K}$ process.}\label{wphikantik}
\end{figure}

Now, the $f_0(1800)$ found in Ref.~\cite{phiwbes} must also be distinct to $f_0(1790)$ of Ref.~\cite{pipibes} since $f_0(1800)$
has been found in the $\phi \omega$ system which must unavoidably couple, and decay, to $K \overline{K}$ (see Fig.~\ref{wphikantik}).
Thus, from Refs.~\cite{phiwbes,pipibes} it seems that two $f_0$ resonances exist with mass around 1800 MeV but with 
very different decay properties and as a consequence there must be three $f_0$'s with mass between 1700-1800 MeV: the well known $f_0(1710)$
and a  twin set of new states around 1800 MeV.  
In this manuscript we discuss that this is not the case although it might appear to be so. We show that the peak structure found in the
$\phi \omega$ spectrum in Ref.~\cite{phiwbes} is a manifestation of $f_0(1710)$ while the resonance seen in the $\pi \pi$ system \cite{pipibes}
is indeed a new scalar, isoscalar resonance which is undoubtedly distinct to $f_0(1710)$. In addition, we provide an explanation for
the suppressed decay of   $f_0(1790)$ to the $K \overline{K}$ channel.

\section{$f_0(1790)$ as a $\pi \pi f_0(980)$ resonance}
We shall first review our study of  the $\pi \pi f_0(980)$ system \cite{uspipi} where a scalar resonance was found  with properties  very similar to $f_0(1790)$.
With the motivation of studying pseudoscalar systems with total quantum numbers of the vacuum,  we first studied three pseudoscalar systems  with total strangeness zero in Ref.~\cite{uspipi}, namely, $\pi^0 K^+ K^-$, $\pi^0 K^0\overline K^0$, $\pi^0\pi^0\eta$, $\pi^+ K^0K^-$, $\pi^+\pi^-\eta$, $\pi^-K^+\overline K^0$ and $\pi^-\pi^+\eta$ as coupled channels. The formalism of the study is based on calculation of the Faddeev equations with the input amplitudes obtained by solving the Bethe-Salpeter equation for the two-pseudoscalar subsystems in a coupled channel formalism. 
The Faddeev partitions, $T^1$, $T^2$ and $T^3$,  are written in this formalism as 
\begin{equation}
T^i =t^i\delta^3(\vec{k}^{\,\prime}_i-\vec{k}_i) + \sum_{j\neq i=1}^3T_R^{ij}, \quad i=1,2,3,
\label{Ti}
\end{equation}
with $\vec{k}_{i}$ ($\vec{k}^\prime_{i}$) being the initial (final) momentum of the particle $i$ and $t^{i}$ the two-body $t$-matrix which describes the interaction of the $(jk)$ pair of the system, $j \neq k\neq i=1,2,3$. Using Eq.~(\ref{Ti}), the full three-body $T$-matrix is obtained in terms of the two-body $t$-matrices and the $T^{ij}_{R}$  partitions as
\begin{equation}
T = T^{1} + T^{2} + T^{3} = \sum_{i=1}^{3}t^i\delta^3(\vec{k}^{\,\prime}_i-\vec{k}_i) +T_{R}\label{T}
\end{equation}
where we define
\begin{equation}
T_{R} \equiv \sum_{i=1}^3\sum_{j\neq i=1}^{3}T^{ij}_{R} . \label{ourfullt}
\end{equation}

The $T^{ij}_{R}$ partitions in Eq.~(\ref{Ti}) satisfy the following set of coupled equations

\begin{equation}
T^{\,ij}_R = t^ig^{ij}t^j+t^i\Big[G^{\,iji\,}T^{\,ji}_R+G^{\,ijk\,}T^{\,jk}_R\Big], \quad i\ne j, j\ne k = 1,2,3. 
  \label{Trest}
\end{equation}
where $g^{ij}$  is a three-body Green's function depending on the variables of the external legs while $G^{ijk}$ is a loop function  (see Refs.~\cite{uspipi,mko1,mko2,mko3,mko4,AJ2} for more details). Studies of several three-hadron systems made of mesons and baryons have earlier been made within this formalism \cite{mko1,mko2,mko3,mko4,AJ2} leading to the generation of some meson and baryon resonances which, in turn, indicates that the  three-body dynamics plays an important role in understanding the properties of such states.

Going back to the study of three pseudoscalar mesons of  Ref.~\cite{uspipi}, we would like to recall that the input two-body amplitudes were obtained following Refs.~\cite{scalar2,scalar3} where dynamical generation of the light scalar mesons was found. Thus, our two-body amplitudes also contain this information. To be precise,  the isoscalar $K\overline{K}$ and $\pi\pi$ $t$-matrices  dynamically generate the resonances $f_0(980)$ and $\sigma(600)$, while the system composed of the channels $K \overline{K}$ and $\pi\eta$ in isospin 1 gives rise to the $a_0(980)$ state. In the strangeness $+1$ $K\pi$ and $K\eta$ systems the $\kappa(850)$ is formed. 

With these inputs we solve  Eq.~(\ref{Trest}) while keeping all the interactions in S-wave, which implies that the total quantum numbers of the three-body system and, thus, the possible bound states or resonances present in it are $J^{\pi}=0^{-}$.  

To identify the peaks obtained in the three-body $T$-matrix for the different channels  with physical states we need to project these amplitudes on an isospin basis. To do that, we consider the total isospin  $I$ of the three-body system and the isospin of one of the two-body subsystems, which in the present case is taken as the isospin of the $K\overline K$ subsystem or (23) subsystem, $I_{23}$, and evaluate the transition amplitude $\langle I,I_{23}|T_R|I,I_{23}\rangle$. The isospin  $I_{23}$ can be 0 or 1, thus, the total isospin $I$ can be 0, 1 or 2. For the cases involving the states $|I=0$, $I_{23}=1\rangle$,  $|I=1$, $I_{23}=1\rangle$ and $|I=2$, $I_{23}=1\rangle$ we have not found any  resonance or bound state. Although we do find one in the case  of $I=1$ with $I_{23}=0$ with mass $~\sim$~1400 MeV and width $\sim$~85 MeV. We find that this resonance is formed when the $K \overline{K}$ system is organized as $f_0(980)$ and, thus, its structure  is dominantly $\pi f_0(980)$.

This state can be associated with the $\pi(1300)$  listed in the PDB \cite{pdg}, whose mass is in the range $1300\pm 100$ MeV and the width found from the different experiments listed varies between 120 to 700 MeV \cite{pdg}. 
Using these values as a reference, the peak position obtained here is in the experimental upper limit  for this state, while the width is close to the lower experimental value, thus,  our findings are compatible with the known data set. Surely, for a better comparison one needs more experiments which could help in determining the properties of this state with more precision. The decay modes seen for this resonance are $\rho \pi$ and $\pi (\pi\pi)_{Swave}$. The channel  $\pi \pi\pi$ is a three-body 
channel which couples to $\pi K\overline K$ and $\pi\pi\eta$. However the three pion threshold
(around 410 MeV) is far away from the region in which the state is formed, thus, it naturally is not essential in the generation of the $\pi(1300)$. However, the inclusion of channels like  $\pi \pi\pi$ or  $\rho \pi$ could help in increasing the width found for the state within our approach, since there is more phase space for the $\pi(1300)$ to decay to these channels.

\begin{figure}[h!]
  \includegraphics[height=5cm]{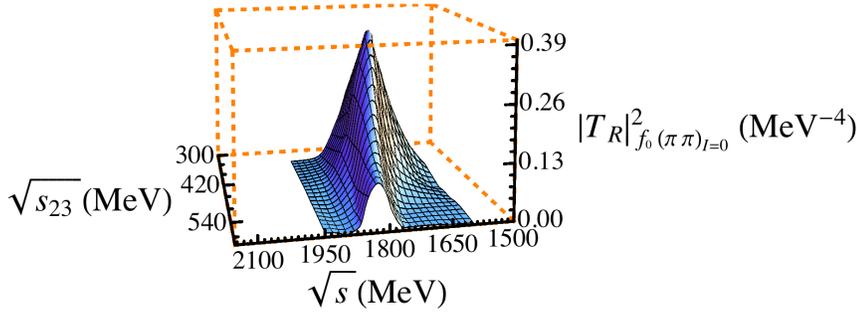}
\caption{The squared amplitude for the $\pi K \overline{K}$ channel as a function of the total energy and the invariant mass of the $K\overline{K}$ system.}\label{f01790}
\end{figure}

\begin{figure}[h!]
  \includegraphics[height=3.4cm]{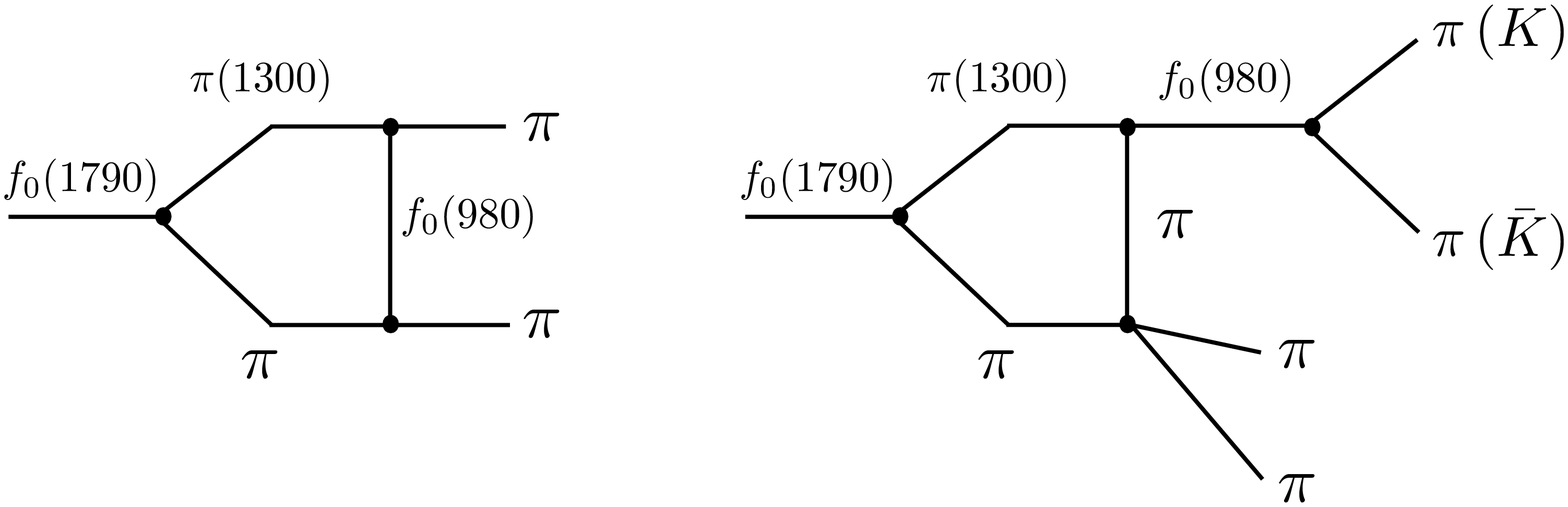}
\caption{The decay process of the $f_0(1790)$  found in our work.}\label{decay}
\end{figure}

Having obtained this information on three pseudoscalar systems we solve the Faddeev equations, once again, for the $\pi \pi f_0(980)$ system. The $\pi \pi$ interaction is obtained from chiral Lagrangians, as earlier, while the result of the Faddeev equations  solved for three-pseudoscalar  system is used for the  $\pi f_0(980)$ amplitude. Consequently, we find a resonance with scalar, isoscalar quantum numbers and mass $\sim$~1773 MeV with 100 MeV width (see Fig.~\ref{f01790}).
 It is found that the two pions interact in the $\sigma(600)$ region and the $\pi f_0(980)$ subsystems are organized as $\pi(1300)$ when the three-body resonance is formed. The important thing to notice from these results is that the scalar, isoscalar resonance found in our work  decays dominantly to $\pi\pi$, $\pi\pi\pi\pi$, $\pi\pi K\overline K$ but not to $K \overline{K}$ (as shown in Fig.~\ref{decay}) which is strikingly similar to the characteristics of the $f_0(1790)$ \cite{pipibes}. 

\section{Manifestation of $f_0(1710)$ in the $\phi \omega$ mass spectrum}
The resonances found in Refs.~\cite{phiwbes,uspipi} can, certainly, not be related to the one found in the $\phi \omega$ invariant mass spectrum in Ref.~\cite{pipibes} since the latter
one must unavoidably decay to $K \overline{K}$ through the mechanism shown in Fig.~\ref{wphikantik}. This argument actually leads to finding of a flaw in the interpretation of the peak seen in the $\phi \omega$ spectrum \cite{phiwbes} as a new $f_0(1800)$ resonance since in the $K\overline K$ decay channel the mass of  $f_0(1800)$  would be very far from the $K\overline K$ threshold and the peak should be clearly observable, with no ambiguities about its interpretation. Yet, in the experiment studying $J/\Psi$ decay into $\gamma K \overline K$, clear peaks are seen for the $f_0(1500)$ and  $f_0(1710)$ but no trace is seen of any peak around 1800 MeV \cite{Bai:2003ww}. Similarly, MARK III~\cite{mark3} reports a clear signal for the $f_0(1710)$ in the $K\overline K$ spectra but no signal around 1800 MeV.

\begin{figure}[h!]
\includegraphics[scale=0.6]{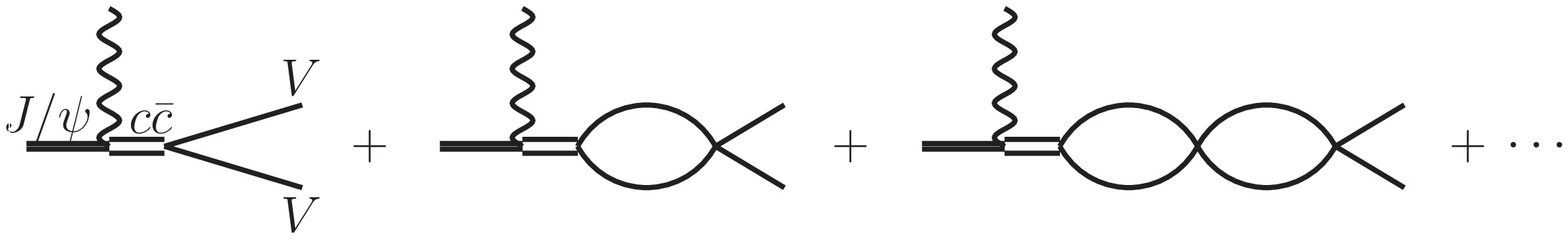}
\caption{Schematic representation of $J/\Psi$ decay into a photon and one dynamically generated resonance.} \label{f3}
\end{figure}

In fact in Ref.~\cite{usphiw} we showed that the peak in the $\phi \omega$ data obtained in Ref.~\cite{phiwbes} can be interpreted as a manifestation of the  $f_0(1710)$ resonance produced below the $\phi \omega$ threshold. We discuss that the presence of this resonance necessarily leads to a peak around the  $\phi \omega$ threshold with a shape and strength compatible with experiment and that the observed peak is not a signal of a new resonance. To do this we studied the $J/\Psi \to \gamma \phi \omega$ process within a formalism where a photon is radiated from the initial $c \overline{c}$ state (as in Refs.~\cite{gengchinos}). The $c \overline{c}$ component after the $\gamma$ radiation, then, decays into pairs of vectors which interact among themselves as shown in Fig.~\ref{f3}. 

The $c \overline{c}$  can be considered as an SU(3) singlet and, thus, the pair of vector mesons  produced after hadronization must couple to an SU(3) singlet. The vector-vector content in the SU(3) singlet can be easily obtained from the trace of $ \mathrm{V\cdot V}$
\begin{equation}
 \mathrm{VV}_{\mbox{SU(3) singlet}}=\mathrm{Tr}[V\cdot V],
\end{equation}
where $V$ is the SU(3) matrix of the vector mesons
\begin{equation}
 V=\left(\begin{array}{ccc}
          \frac{1}{\sqrt{2}}\rho^0+\frac{1}{\sqrt{2}}\omega & \rho^+ &K^{*+}\\
           \rho^- &-\frac{1}{\sqrt{2}}\rho^0+\frac{1}{\sqrt{2}}\omega & K^{*0}\\
           K^{*-} &\overline{K}^{*0}&\phi
         \end{array}
\right).
\end{equation}
We, thus, find the vertex
\begin{equation}
 \mathrm{VV}_{\mbox{SU(3) singlet}}=
\rho^0\rho^0+\rho^+\rho^-+\rho^-\rho^++\omega\omega+K^{*+}K^{*-}+K^{*0}\overline{K}^{*0}
+K^{*-}K^{*+}+\overline{K}^{*0}K^{*0}+\phi\phi.
\end{equation}

It is important to note that there is no primary production of $\phi \omega$ with the mechanism of Fig.~\ref{f3}. The production of $\phi \omega$ occurs through the rescattering of the  vector mesons  produced primarily. The amplitude for the process shown in Fig.~\ref{f3} can be written as
\begin{equation}
t_{J/\Psi \rightarrow \gamma \phi \omega} = A \sum\limits_{j=1}^4 w_j G_j t_{j\rightarrow \phi \omega},\label{eq:tjpsi}
\end{equation}
where $A$ is an unknown constant representing the reduced matrix element for the operator responsible for the transition $c\overline c\to \textrm{VV}_{\textrm{SU(3) singlet}}$, $w_j$ are the weight factors corresponding to the probability of hadronization of  different vector mesons, $G_j$ the loop function for the intermediate two mesons state and  $t_{j\rightarrow \phi \omega}$ represents the transition amplitude for  the intermediate vector mesons to $\phi \omega$. We take the information for the $G_j$ and $t_{ij}$ functions from Ref.~\cite{gengvec}.  
 The $t_{i\to j}$ matrices can be written as
\begin{equation}
t_{i\to j} = \frac{g_i g_j}{s- M^{2}_R + i M_R\Gamma_{R}} \label{eq:bw}
\end{equation}
where $g_i, g_j$ are the couplings of the resonance to the $i, j$ channels given in  Ref.~\cite{gengvec}.  

With the amplitude of Eq.~(\ref{eq:tjpsi}), which depends on the invariant mass of $\phi \omega$, we can construct the $\phi \omega$ mass distribution given by
\begin{equation}
\frac{d\Gamma}{dM_{\rm inv}} = \frac{1}{(2\pi)^3} \frac{1}{ 4 M_{J/\Psi}^2} \,\,p_{\gamma} \overline{q}_{\omega} \mid t_{J/\Psi \rightarrow \gamma \phi \omega} \mid^2, \label{eq:totalgamma}
\end{equation}
where $p_{\gamma}$ and $\overline{q}_{\omega}$ are the photon momentum in the $J/\Psi$ rest frame and the $\omega$ momentum in the $\phi \omega$ rest frame, respectively.

\begin{figure}[htbp]
\includegraphics[scale=0.26]{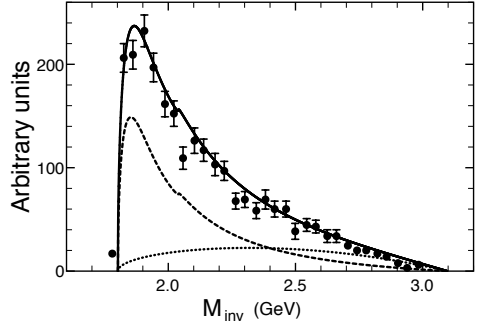}
\caption{The invariant mass distribution $\frac{d\Gamma}{d M_{inv}}$ for the process $J/\Psi \rightarrow \gamma \phi \omega$ from Eq.~(\ref{eq:totalgamma}). The data points, shown by filled circles, have been taken from Ref.~\cite{expnew}. The dotted and dashed lines represent the background  and the $f_0(1710)$ resonance contribution, respectively. The solid line shows the coherent sum of the two. } \label{invmass}
\end{figure}

In Fig.~\ref{invmass} we show the $\phi\omega$ invariant mass distribution obtained by fixing the total strength such as to reproduce
the peak of the experimental data on the number of $\phi \omega$ events per bin. In order to account for the strength of the distribution at large values of $M_{\rm inv}$, far away from the $f_0 (1710)$ resonance, we allow for a small background, which we take as a constant amplitude for simplicity. As we can see, there is a perfect agreement between our results and the experimental data. However, we must remember that we have an unknown constant $A$ in Eq.~\ref{eq:tjpsi}. In order to make a stronger claim, we have calculated the ratio 
\begin{equation}
R_{\Gamma} = \frac{\displaystyle\int d M_{\rm inv} \frac{d\Gamma}{dM_{\rm inv}}}{\Gamma_{J/\Psi \rightarrow \gamma f_0 (1710)}}, \label{eq:ratio}
\end{equation}
where the constant $A$ gets cancelled. In our work this ratio  turns out to be $0.15^{+ 0.07}_{- 0.04}$ which is in good agreement with the experimental value $0.14^{+ 0.12}_{- 0.07}$.

With this we can summarize the present manuscript by mentioning that  we provide evidence for existence of a new scalar, isoscalar resonance $f_0(1790)$ which is distinct to the known $f_0(1710)$. We present an explanation for the suppressed decay of this resonance to $K \overline{K}$. We also show that the BES data on the $\phi \omega$ spectrum can be explained in terms of $f_0(1710)$ and thus a new $f_0(1800)$ is not required. Thus there are two $f_0$ states in the 1700-1800 MeV.

\begin{theacknowledgments}
The authors acknowledge the support from the funding agencies FAPESP and CNPq.
\end{theacknowledgments}



\bibliographystyle{aipproc}   

\bibliography{sample}

\IfFileExists{\jobname.bbl}{}
 {\typeout{}
  \typeout{******************************************}
  \typeout{** Please run "bibtex \jobname" to optain}
  \typeout{** the bibliography and then re-run LaTeX}
  \typeout{** twice to fix the references!}
  \typeout{******************************************}
  \typeout{}
 }


\end{document}